\documentstyle[12pt,epsf]{article}

\begin{document}
\title{
\vspace*{-4ex} \hfill {\large UT-733 \ \ \ } \\
  \vspace{5ex}
  Quantum Restoration of the U(1)$_{Y}$ Symmetry \\
  in Dynamically Broken SUSY-GUT's
}
\author{T. Hotta, Izawa K.-I. and T. Yanagida\\
  \\  Department of Physics, University of Tokyo \\
  Bunkyo-ku, Tokyo 113, Japan}
\date{\today}
\maketitle
\setlength{\baselineskip}{3.6ex}
\begin{abstract}
  We propose a supersymmetric hypercolor SU(3)$_{H}$ gauge theory
  interacting strongly at the grand unification scale, in which the
  hyperquark condensation breaks SU(5)$_{GUT}$ down to SU(3)$_{C}$
  $\times$ SU(2)$_{L}$ without unbroken U(1)$_{Y}$ at the classical
  level.
  However, we show that the broken U(1)$_{Y}$ symmetry is restored by
  quantum mechanical effects and hence there remains the
  standard-model gauge symmetry at the electroweak scale.
  The dynamics of the strong interactions also produces naturally a
  pair of massless Higgs doublets.
  In addition to these Higgs doublets, we have a pair of massless
  singlets which contributes to the renormalization-group equations of
  gauge coupling constants and hence affects the GUT unification.
  We discuss a simple solution to this problem.
\end{abstract}

\newpage

\section{Introduction}

An SU(2)$_{L}$ \hspace{-0.5em}-doublet Higgs boson of mass of the
order of hundred GeV is an inevitable ingredient in the standard
electroweak gauge model.
In the grand unified theories (GUT's) \cite{gut} one must require an
extremely precise fine-tuning among various parameters in order to
obtain the light doublet Higgs boson.
The situation is not improved much even if one assumes supersymmetry
(SUSY) \cite{susygut}.
Therefore, the necessity of the fine-tuning of parameters is regarded
as a crucial drawback in the GUT scenario, although the recent
high-precision measurements on the gauge coupling constants have
strongly supported the SUSY extension of the GUT's \cite{preciceexp}.

In recent papers \cite{our1,our2,hisano} we have proposed a SUSY gauge
theory based on SU(3)$_{H}$ $\times$ U(1)$_{H}$ with six flavors of
quarks which interact strongly at the GUT scale and shown that the
dynamics of this theory produces  a pair of massless Higgs doublets.
Thus, we have a natural explanation for the presence of light Higgs
doublets at the electroweak scale in the SUSY-GUT's.

However, we have also found that the hyperquark condensation breaks
the GUT gauge group SU(5)$_{GUT}$ down to SU(3)$_{C}$ $\times$
SU(2)$_{L}$ \hspace{-0.3em}.
Therefore, we need to introduce an extra U(1)$_{H}$ to have the
standard-model gauge group unbroken at the low energies
\cite{our1,our2,hisano}.
The presence of U(1)$_{H}$ above the GUT scale is nothing wrong
phenomenologically, but it would be more interesting if we did not
need to introduce the extra U(1)$_{H}$ \hspace{-0.3em}.

In this paper we show that introduction of a singlet field $\Phi$ into
the previous model solves the problem.
Namely, in this new model, the broken U(1)$_{Y}$ (a subgroup of
SU(5)$_{GUT}$) is restored by quantum mechanical effects and hence we
do not need to introduce the extra U(1)$_{H}$ \hspace{-0.3em}.
In fact, we find a stable quantum vacuum corresponding to the desired
symmetry breaking SU(5)$_{GUT}$ $\rightarrow$ SU(3)$_{C}$ $\times$ %
SU(2)$_{L}$ $\times$ U(1)$_{Y}$ \hspace{-0.3em}.
We also show that there remains a pair of Higgs doublets massless in
this quantum vacuum.
In addition to these Higgs doublets, however, we have a pair of
massless singlets which contributes to the renormalization-group flows
of low-energy gauge coupling constants and affects the unification of
gauge couplings at the GUT scale.
We also propose a simple solution to this problem, giving a
sufficiently large mass to the singlet pair.

\section{The model}

Let us consider a supersymmetric hypercolor SU(3)$_{H}$ gauge theory
\cite{our1,our2} with 6 flavors of hyperquark chiral superfields
$Q_\alpha^A$ and $\bar{Q}^\alpha_A$ $(\alpha = 1, \cdots ,3)$ in the
fundamental representations {\bf 3} and $\bf 3^*$ of SU(3)$_{H}$ ,
respectively.
Here, $A$ denotes flavor index and it runs from 1 to 6.
The first five $Q_\alpha^I$ and $\bar{Q}^\alpha_I$
$(I = 1 , \cdots, 5)$ transform as $\bf 5^*$ and {\bf 5},
respectively, under the GUT gauge group SU(5)$_{GUT}$, while the last
$Q_\alpha^6$ and $\bar{Q}^\alpha_6$ are singlets of SU(5)$_{GUT}$
\cite{our2}.

We introduce three kinds of SU(3)$_{H}$ \hspace{-0.8ex}-singlet chiral
superfields, a pair of $H_I$ and $\bar{H}^I$, $\Sigma^I{}_{\! J}$ and
$\Phi$ $(I, J = 1,\cdots, 5)$, which are {\bf 5}, $\bf 5^*$, {\bf 24}
$\! + \!$ {\bf 1} and {\bf 1} of SU(5)$_{GUT}$ \hspace{-0.5ex}.
Besides the gauge symmetry, SU(3)$_{H}$ $\times$ SU(5)$_{GUT}$
\hspace{-0.5ex}, we impose a global U(1)$_{A}$ symmetry
\cite{our1,our2}
\begin{equation}
\begin{array}{ccl}
  Q_\alpha^I, \bar{Q}^\alpha_I & \rightarrow & Q_\alpha^I,
  \bar{Q}^\alpha_I , \\
  \noalign{\vskip 1ex}
  Q_\alpha^6, \bar{Q}^\alpha_6 & \rightarrow & e^{i \xi} Q_\alpha^6,
  e^{i \xi} \bar{Q}^\alpha_6 , \\
  \noalign{\vskip 1ex}
  H_I, \bar{H}^I & \rightarrow & e^{- i \xi} H_I, e^{- i \xi}
  \bar{H}^I , \\
  \noalign{\vskip 0.5ex}
  \Sigma^I{}_{\! J} & \rightarrow & \Sigma^I{}_{\! J} , \\
  \noalign{\vskip 0.5ex}
  \Phi & \rightarrow & e^{-2 i \xi} \Phi , \\
  \noalign{\vskip 0.3ex}
  & & \hspace*{-4em} (I = 1, \cdots, 5).
\end{array}
\end{equation}

With the above gauge and global symmetries, we have a superpotential
\begin{equation}
\begin{array}{rl}
  \label{superpotential}
  W = & \lambda \Sigma^I{}_{\! J} \, \bar{Q}^\alpha_I \, Q^J_\alpha
      + h H_I \bar{Q}^\alpha_6 Q^I_\alpha
      + h' \bar{H}^I \bar{Q}^\alpha_I Q_\alpha^6
      + f \Phi \, \bar{Q}^\alpha_6 Q_\alpha^6 \\
      \noalign{\vskip 1ex}
      & \displaystyle + \frac{1}{2} \, m_\Sigma Tr(\Sigma^2)
      + \frac{1}{2} \, m'_\Sigma (Tr \Sigma)^2 - \mu_\Sigma Tr \Sigma
      \, .
\end{array}
\end{equation}
Here, we have omitted trilinear self-coupling terms, $Tr(\Sigma^3)$,
$Tr(\Sigma^2) Tr \Sigma \,$, and $(Tr\Sigma)^3$, for simplicity, and
the mass term, $m_Q \bar{Q}^\alpha_I \, Q_\alpha^I $, has been
absorbed into the $\Sigma$ field by shift of the trace part of
$\Sigma$
\footnote{One may assume the hyperquark mass term,
  $m_Q \bar{Q}^\alpha_I Q^I_\alpha$, instead of introducing the trace
  part of $\Sigma$. For this case, one may also derive the same
  conclusion in the present paper.}.
The terms, $\bar{H}^I H_I$, $\bar{Q}^\alpha_6 Q_\alpha^6$ and
$\Phi^n (n = 1, \cdots, 3)$, for example, are forbidden by the global
U(1)$_{A}$ \hspace{-0.5ex}.
Notice that the singlet $\Phi$ is not assumed in the previous models
\cite{our1,our2}.
As we will see later, $\Phi$ plays a crucial role on quantum
mechanical restoration of the U(1)$_{Y}$ gauge symmetry in the
present model.

We first consider the following classical vacuum discussed in the
previous analyses \cite{our1,our2}:
\begin{equation}
  \label{classicalvacuum}
  \begin{array}{l}
    \langle Q_\alpha^A \rangle = \left(
    \begin{array}{ccc}
      0 & 0 & 0 \\
      0 & 0 & 0 \\
      v & 0 & 0 \\
      0 & v & 0 \\
      0 & 0 & v \\
      0 & 0 & 0
    \end{array}
  \right),\
  \langle \bar{Q}^\alpha_A \rangle = \left(
  \begin{array}{cccccc}
    0 & 0 & v & 0 & 0 & 0 \\
    0 & 0 & 0 & v & 0 & 0 \\
    0 & 0 & 0 & 0 & v & 0
  \end{array}
\right), \\
\noalign{\vskip 2ex}
\displaystyle \langle \Sigma^I{}_{\! J} \rangle =
\frac{\mu_\Sigma}{m_\Sigma + 2 m'_\Sigma}
\left(
  \begin{array}{ccccc}
    1 & & & & \\
    & 1 & & & \\
    & & 0 & & \\
    & & & 0 & \\
    & & & & 0
  \end{array}
\right) \, , \\
\noalign{\vskip 1ex}
\langle H_I \rangle = \langle \bar{H}^I \rangle = 0 \, ,
\end{array}
\end{equation}
with $A = 1, \cdots, 6$; $\, I, J = 1, \cdots ,5$ and
\begin{equation}
v = \sqrt{\frac{m_\Sigma \, \mu_\Sigma}{\lambda(m_\Sigma + 2
    m'_\Sigma)}} \, .
\end{equation}
Here, the vacuum-expectation value of $\Phi$ is undetermined since its
potential is flat for $\langle Q_\alpha^6 \rangle = \langle
\bar{Q}^\alpha_6 \rangle = 0$.
In this classical vacuum the gauge group is broken down as
\begin{equation}
  SU(3)_H \times SU(5)_{GUT} \rightarrow SU(3)_C \times SU(2)_L \, .
\end{equation}
There is no unbroken U(1)$_{Y}$ \hspace{-.5em}, which is the main
motivation to introduce an extra U(1)$_{H}$ gauge symmetry in
Ref.\cite{our1,our2}.
We now show that U(1)$_{Y}$ broken in the classical vacuum is restored in
a quantum vacuum.

\section{Restoration of the U(1)$_{Y}$ symmetry}

Let us investigate the quantum vacua which satisfy the following
conditions:
(i) the vacuum-expectation values of $\Sigma$ have a diagonal form
with those of $H$ and $\bar H$ vanishing, and
(ii) at least two hyperquarks among $Q_\alpha^I$ and
$\bar{Q}^\alpha_I$ $(I = 1, \cdots, 5)$ are massive in the vacua so
that we can integrate out them to get a SUSY QCD-like theory with the
effective $N_f = 4$ flavors.
We identify the two massive hyperquarks with $Q^I_\alpha$ and
$\bar{Q}^\alpha_I$ $(I = 1, 2)$.
These quantum vacua include the vacuum defined in
Eq.(\ref{classicalvacuum}) in the classical limit.

We analyze the vacua in a similar way taken in Ref.\cite{our2}.
The integration of massive hyperquarks, $Q^I_\alpha$ and
$\bar{Q}^\alpha_I$ $(I = 1, 2)$, leads to a low-energy effective
theory with a superpotential
\begin{equation}
\begin{array}{rl}
  W_{low} = & \lambda \Sigma^a{}_{\! b} \bar{Q}^\alpha_a \, Q^b_\alpha
        + h H_a \bar{Q}^\alpha_6 Q^a_\alpha
        + h' \bar{H}^a \bar{Q}^\alpha_a Q_\alpha^6 \\
        \noalign{\vskip 1ex}
        & \displaystyle + f \Phi \, \bar{Q}^\alpha_6 Q_\alpha^6
        - h h' \left( \frac{\bar{H}^1 H_1}{m_1}
        + \frac{\bar{H}^2 H_2}{m_2} \right) \bar{Q}^\alpha_6 \,
        Q_\alpha^6 \\
        & \displaystyle + \frac{1}{2} \, m_\Sigma \Sigma^a{}_{\! b}
        \Sigma^b{}_{\! a}
        + \frac{1}{2} \, {m_\Sigma m'_\Sigma \over m_\Sigma + 2m'_\Sigma}
        ( \Sigma^a{}_{\! a} )^2
        - {m_\Sigma \mu_\Sigma \over m_\Sigma + 2m'_\Sigma}
        \Sigma^a{}_{\! a} \, ,
\end{array}
\end{equation}
where $a, b = 3, 4, 5$, and $m_1$ and $m_2$ are the masses for the
first and second hyperquarks, respectively
\footnote{We have assumed $\langle Q_\alpha^1 \bar{Q}^\alpha_1 \rangle
  = \langle Q_\alpha^2 \bar{Q}^\alpha_2 \rangle = 0$, which is
  confirmed in view of the following analysis.}.
Here, $\Sigma^I{}_{\! J}$ $(I, J = 1, 2)$ are decoupled from this
low-energy superpotential and we have omitted $\Sigma^I{}_{\! i}$ and
$\Sigma^i{}_{\! I}$ $(I = 1, 2, \; i = 3, \cdots, 6)$ since they are
irrelevant for our analysis.

The effective superpotential which incorporates the strong SU(3)$_{H}$
dynamics is described by meson $M^i{}_{\! j}$ and baryon $B_i$,
$\bar{B}^i$ chiral superfields
\cite{seiberg1}
\begin{eqnarray}
  \begin{array}{lll}
    & & M^i{}_{\! j} \sim Q^i_\alpha \bar{Q}^\alpha_j \, , \\
    \noalign{\vskip 1ex}
    & & \displaystyle B_i \sim \epsilon^{\alpha \beta \gamma}
    \epsilon_{i j k l} Q^j_\alpha Q^k_\beta Q^l_\gamma \, , \\
    \noalign{\vskip 1ex}
    & & \displaystyle \bar{B}^i \sim \epsilon_{\alpha \beta \gamma}
    \epsilon^{i j k l} \bar{Q}_j^\alpha \bar{Q}_k^\beta
    \bar{Q}_l^\gamma \, , \\
    \noalign{\vskip 1ex}
  \end{array}
\end{eqnarray}
as follows:
\begin{equation}
\label{weff}
\begin{array}{rl}
  W_{eff} = & \Lambda^{-5} ( B_i M^i{}_{\! j} \bar{B}^j
            - \det M^i{}_{\! j} )
            + \lambda \Sigma^a{}_{\! b} M^b{}_{\! a} \\
            \noalign{\vskip 1ex}
            & + h H_a M^a{}_{\! 6} + h' \bar{H}^a M^6{}_{\! a}
            + f \Phi \, M^6{}_{\! 6} \\
            \noalign{\vskip 1ex}
            & \displaystyle - h h' \left( \frac{\bar{H}^1 H_1}{m_1}
            + \frac{\bar{H}^2 H_2}{m_2} \right) M^6{}_{\! 6} \\
            \noalign{\vskip 1ex}
            & \displaystyle + \frac{1}{2} \, m_\Sigma \Sigma^a{}_{\! b}
            \Sigma^b{}_{\! a}
            + \frac{1}{2} \, {m_\Sigma m'_\Sigma \over m_\Sigma +
              2m'_\Sigma} ( \Sigma^a{}_{\! a} )^2
            - {m_\Sigma \mu_\Sigma \over m_\Sigma + 2m'_\Sigma}
            \Sigma^a{}_{\! a} \, ,
\end{array}
\end{equation}
where $\Lambda$ denotes a dynamical scale of the low-energy
SU(3)$_{H}$ interactions, $a, b = 3, \cdots, 5$, and $i, j, k, l = 3,
\cdots, 6$
\footnote{The global U(1)$_{A}$ has, of course, SU(3)$_{H}$ gauge
  anomalies and it is broken by the nonperturbative effects.
  However, the effective superpotential $W_{eff}$ in Eq.(\ref{weff})
  except the dynamical part
  $(B_i M^i{}_{\! j} \bar{B}^j - \det M^i{}_{\! j})$ respects the
  U(1)$_{A}$ symmetry as noted in Ref.\cite{our2}.}.

For $\langle \Phi \rangle = 0$, we can see that $B_6$ and $\bar{B}^6$
necessarily have non-vanishing vacuum-expectation values leading to
the breaking of U(1)$_{Y}$ as in the classical vacuum in
Eq.(\ref{classicalvacuum}).
However, there is a desired supersymmetric vacuum where
\begin{equation}
\begin{array}{l}
  \label{quantumvacuum}
  \langle B_i \rangle = \langle \bar{B}^i \rangle = 0 \, ,\\
  \noalign{\vskip 0.5ex}
  \langle M^6{}_{\! 6} \rangle = \langle M^6{}_{\! a} \rangle =
  \langle M^a{}_{\! 6} \rangle = 0 \, ,\\
  \noalign{\vskip 0.5ex}
  \langle \Sigma^a{}_{\! b} \rangle = 0 \, ,\\
  \displaystyle \langle M^a{}_{\! b} \rangle =
  {m_\Sigma \mu_\Sigma \over \lambda (m_\Sigma + 2m'_\Sigma)} \,
  \delta^a{}_{\! b} \, , \quad
  \langle \Phi \rangle = \frac{1}{f \Lambda^5}
  \left[ {m_\Sigma \mu_\Sigma \over \lambda
  (m_\Sigma + 2m'_\Sigma)} \right]^3 \, , \\
  \noalign{\vskip 0.5ex}
  (i = 3, \cdots, 6 \; {\rm and} \; a, b = 3, \cdots, 5) \, .
\end{array}
\end{equation}
Notice that in this vacuum the sixth hyperquark has an effective mass
$m_6 = f \langle \Phi \rangle$ while other three hyperquarks are
massless.
Therefore, our model becomes a SUSY QCD-like theory with $N_f = 3$
massless quarks.
The quantum vacuum in Eq.(\ref{quantumvacuum}) is not the same as the
classical one, which is consistent with the result obtained by Seiberg
\cite{seiberg1} for the case of $N_f = N_c$, with $N_c$ being the
number of colors.

With $\langle M^6{}_{\! 6} \rangle = 0$ we find that the Higgs
doublets $H_I$ and $\bar{H}^I$ $(I = 1, 2)$ remain massless in this
quantum vacuum \cite{our2}.
On the other hand, their color partners $H_a$ and $\bar{H}^a$ have
GUT-scale masses.

When $\langle \Phi \rangle \neq 0$ as in Eq.(\ref{quantumvacuum}), the
sixth hyperquark becomes massive.
Therefore, the above vacuum in Eq.(\ref{quantumvacuum}) may be also
studied by means of the effective superpotential obtained through
integrating out the sixth hyperquark instead of the second one:
\begin{equation}
\begin{array}{rl}
  W'_{eff} = & \Lambda'^{-5}
             ( B_i M^i{}_{\! j} \bar{B}^j - \det M^i{}_{\! j} )
             + \lambda \Sigma^i{}_{\! j} M^j{}_{\! i} \\
             \noalign{\vskip 1ex}
             & \displaystyle
             - \frac{h h'}{m_6} H_i \bar{H}^j M^i{}_{\! j}
             + \frac{1}{2} \, m_\Sigma \Sigma^i{}_{\! j} \Sigma^j{}_{\! i}
             + \frac{1}{2} \, {m_\Sigma m'_\Sigma \over m_\Sigma +
               m'_\Sigma}
             ( \Sigma^i{}_{\! i} )^2
             - {m_\Sigma \mu_\Sigma \over m_\Sigma + m'_\Sigma}
             \Sigma^i{}_{\! i} \, ,
\end{array}
\end{equation}
where $i, j = 2, \cdots, 5$.
The dynamical scale $\Lambda'$ is different from the previous
$\Lambda$ by a simple-threshold relation \cite{int} $m_2 \Lambda'^5 =
m_6 \Lambda^5$.
{}From this superpotential we obtain
\begin{equation}
  \langle M^2{}_{\! 2} \rangle = 0, \quad
  \langle \Sigma^2{}_{\! 2} \rangle
  = {\mu_\Sigma \over m_\Sigma + 2m'_\Sigma} \, .
\end{equation}
Similarly we obtain
$\langle M^1{}_{\! 1} \rangle = 0$ and
$\langle \Sigma^1{}_{\! 1} \rangle = \langle \Sigma^2{}_{\! 2} \rangle$,
which lead us to conclude that in the quantum vacuum in
Eq.(\ref{quantumvacuum}) we have the desired breaking of the GUT gauge
group down to the standard-model one $SU(3)_C \times SU(2)_L \times
U(1)_Y$.

As mentioned in the introduction, we have a pair of massless bound
states $B_6$ and $\bar{B}^6$.
Since they have U(1)$_{Y}$ \hspace{-1ex}-charges, they contribute to
the renormalization-group equations of three gauge coupling constants
in the standard model.
We now show that a natural extension of our original superpotential
given in Eq.(\ref{superpotential}) induces a sufficiently large mass
to this dangerous singlet pair.

It seems quite reasonable that there are nonrenormalizable operators
in the superpotential suppressed by some scale $M_0$ higher than the
GUT scale (originating from gravitational interaction, for example).
We, thus, consider the lowest-dimensional nonrenormalizable operator
consistent with our gauge and global symmetries that is to contain
baryon superfields.
That is
\begin{equation}
  \label{highdimop}
  \delta W = \frac{f'}{M^3_0} \epsilon^{\alpha \beta \gamma}
  \epsilon_{\alpha' \beta' \gamma'} (Q^I_\alpha Q^J_\beta Q^K_\gamma)
  (\bar{Q}^{\alpha'}_I \bar{Q}^{\beta'}_J \bar{Q}^{\gamma'}_K) \, .
\end{equation}
This interaction generates a mass term for $B_6$ and $\bar{B}^6$ in
the effective superpotential as
\begin{equation}
  \delta W_{eff} = \frac{f'}{M^3_0} B_6 \bar{B}^6 \, ,
\end{equation}
which corresponds to the physical mass for $B_6$ and $\bar{B}^6$
\begin{equation}
  m_{B_6} \simeq \frac{f' \Lambda^4}{M^3_0} \, .
\end{equation}
Taking $f' \sim {\cal O}(1)$,  $\Lambda \sim 10^{16}-10^{17}$ GeV and
$M_0 \sim 10^{17}-10^{18}$ GeV, we obtain $m_{B_6} \sim
10^{13}-10^{14}$ GeV.
In Fig.1 we show two-loop renormalization-group flows of three gauge
coupling constants, $\alpha_1, \alpha_2$ and $\alpha_3$, with $m_{B_6}
= 10^{13}$ GeV and $\alpha_3(m_Z) = 0.117 \pm 0.010 (2\sigma)$
\cite{PDG}.
We see that our model is marginally consistent with the GUT
unification and the small value of $\alpha_3$ is favored.
It may be interesting that the recently reported $Zbb$ anomaly implies
rather small $\alpha(m_Z)$ \cite{hagiwara}.

It is remarkable that the higher-dimensional operator in
Eq.(\ref{highdimop}) stabilizes the present vacuum given in
Eq.(\ref{quantumvacuum}) and there is no flat direction around the
vacuum besides the directions of $\langle H_I \rangle = \langle
\bar{H}^I \rangle$ $(I = 1, 2)$
\footnote{Without the new term in Eq.(\ref{highdimop}) we have another
    flat direction given by $\langle B_6 \rangle \langle \bar{B}^6
  \rangle + f \langle \Phi \rangle = \det \langle M^a{}_{\! b} \rangle
  $.}.

\section{Conclusion}

In this paper we have proposed a model for the dynamical generation of
light Higgs doublets in SUSY-GUT's based on a semisimple gauge group
SU(3)$_{H}$ $\! \! \times \!$ SU(5)$_{GUT}$ \hspace{-0.7ex}.
The SU(3)$_{H}$ interactions become strong at the GUT scale and cause
a dynamical breaking of the SU(5)$_{GUT}$ \hspace{-0.3em}.
We have found that there is indeed a desired quantum vacuum
corresponding to the breaking $SU(5)_{GUT} \rightarrow SU(3)_C \times
SU(2)_L \times U(1)_Y$.
We have also noted that an introduction of a higher-dimensional
operator in Eq.(\ref{highdimop}) to the superpotential stabilizes the
desired quantum vacuum.
Moreover, the same operator gives rise to a mass for the composite
baryon states $B_6$ and $\bar{B}^6$ of the order of $10^{13}-10^{14}$
GeV, which makes our model consistent with the GUT unification of
three gauge coupling constants of $SU(3)_C \times SU(2)_L \times
U(1)_Y$.

Finally, we comment on the proton decay via the dimension-five
operators \cite{dimfiveop}.
In the present vacuum, the mass matrix for color-triplet states $M, H$
and $\bar{H}$ is given by
\begin{equation}
  ( H_a \; M^6{}_{\! a} ) \; \hat{m}_C
  \left( \begin{array}{c}
    \bar{H}^a \\ M^a{}_{\! 6}
  \end{array} \right),
\end{equation}
where
\begin{equation}
  \hat{m}_C =
  \left( \begin{array}{cc}
    \displaystyle 0 & h \\
    \noalign{\vskip 0.5ex}
    h' & \displaystyle \frac{v^4}{\Lambda^5}  \\
  \end{array} \right).
\end{equation}
with
$\displaystyle v^2 =
\frac{m_\Sigma \mu_\Sigma}{\lambda(m_\Sigma + 2m'_\Sigma)}$.
The dimension-five operators are proportional to
$\left( \hat{m}_C^{-1} \right)_{11}$, which is at the order of the GUT
scale, $\displaystyle \sim \frac{v^4}{h h' \Lambda^5}$.
Thus, we have unsuppressed proton decays
\footnote{From the present experimental limit on the proton decays we
  derive a constraint \cite{hisano2} $h h' \Lambda^5 / v^4
  \stackrel{>}{\mbox{\raisebox{-0.6ex}{$\sim$}}} 3 \times 10^{16}$ GeV.}.
This is a crucial point to distinguish the present model from the
previous one \cite{our1,our2}.
The reason for having non-vanishing dimension-five operators is that
$\langle \Phi \rangle \neq 0$ induces the effective mass for
$Q^6_\alpha$ and $\bar{Q}^\alpha_6$ and hence there remains no
U(1)$_{A}$ symmetry
\footnote{Notice that $\langle \Phi \rangle \neq 0$ never generates
  the mass term for $H_I$ and $\bar{H}^I (I = 1, 2)$ as seen in
  Eq.(\ref{weff}).}
, which forbids the dimension-five operators for the proton decays.

\newpage
\section*{\large Figure caption }
\renewcommand{\labelenumi}{Fig.\arabic{enumi}}
\begin{enumerate}
\item Two-loop renormalization-group flows of three gauge coupling
  constants, $\alpha_1$, $\alpha_2$ and $\alpha_3$, for $m_{B_6} =
  10^{13}$ GeV.
  The solid (dashed) lines are obtained for $\alpha_3(m_Z) = 0.107
  \,(0.127)$ which are within $2 \sigma$ ranges from mean value
  \cite{PDG}.
  We have assumed the SUSY-breaking scale to be 1 TeV.
\end{enumerate}

\newpage
\pagestyle{empty}
\begin{figure}[tbp]
  \hspace*{-10ex}
  \epsfbox{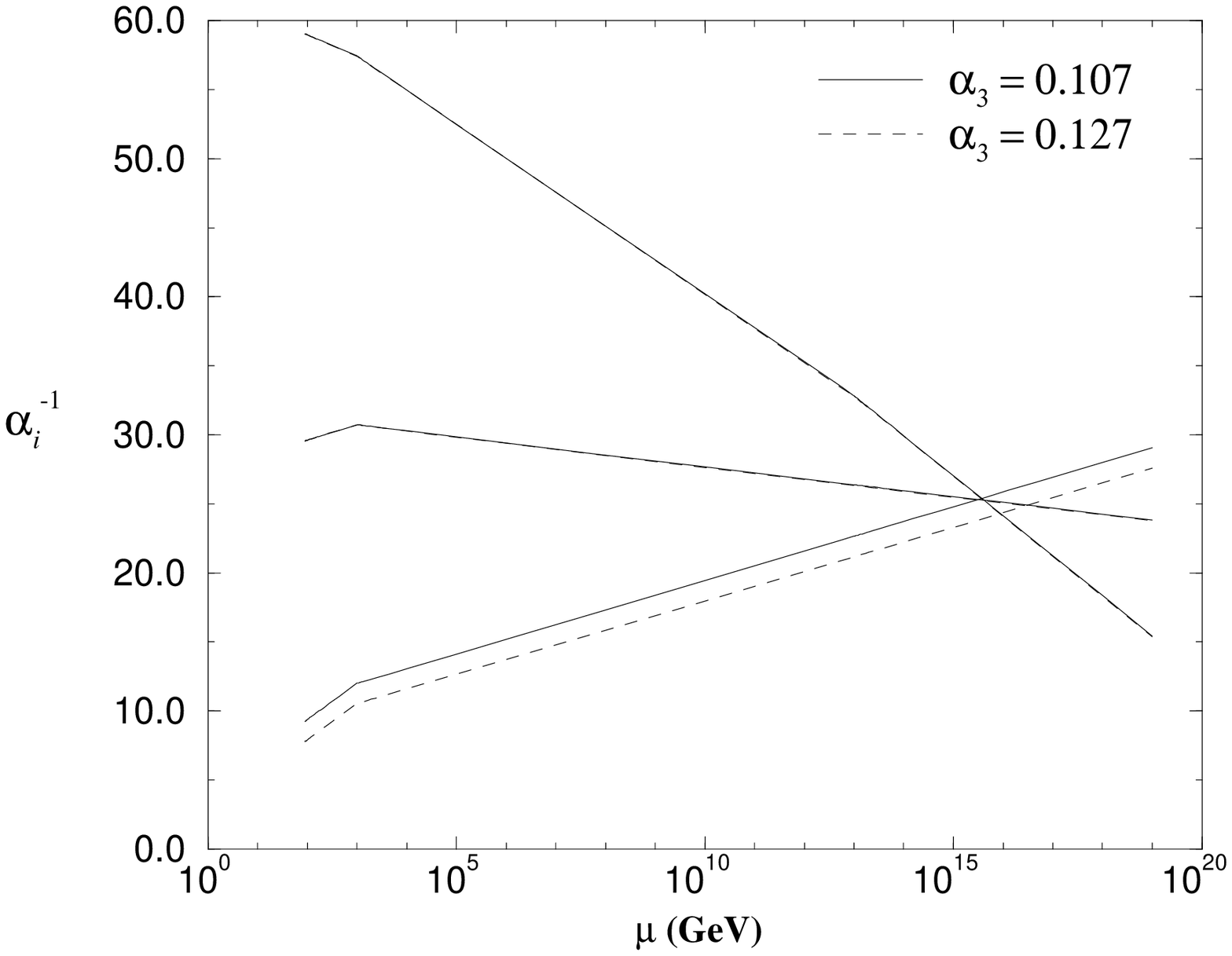}
\end{figure}


\newpage

\begin{thebibliography}{99}
\bibitem{gut} H.~Georgi and S.L.~Glashow, Phys. Rev. Lett. {\bf 32},
  438 (1974).

\bibitem{susygut} E.~Witten, Nucl. Phys. {\bf B188}, 513 (1981); \\
  S.~Dimopoulos, S.~Raby and F.~Wilczek, Phys. Rev. {\bf D24}, 1681
  (1981); \\
  S.~Dimopoulos and H.~Georgi, Nucl. Phys. {\bf B193}, 150 (1981); \\
  N.~Sakai, Z. Phys. {\bf C11}, 153 (1981).

\bibitem{preciceexp}
  P.~Langacker and M.-X.~Luo, Phys. Rev. {\bf D44}, 817 (1991); \\
  U.~Amaldi, W.~de~Boer and H.~F\"urstenau, Phys. Lett. {\bf B260}, 447
  (1991); \\
  J.~Ellis, S.~Kelley and D.V.~Nanopoulos, Phys. Lett. {\bf B260}, 131
  (1991); \\
  W.J.~Marciano, Brookhaven preprint BNL-45999 (April 1991).

\bibitem{our1} T.~Yanagida, Phys. Lett. {\bf B344}, 211 (1995).

\bibitem{our2} T. Hotta, Izawa K.-I. and T. Yanagida, hep-ph/9509201.

\bibitem{hisano} J.~Hisano and T.~Yanagida, hep-ph/9510277.

\bibitem{seiberg1} N.~Seiberg, Phys. Rev. {\bf D49}, 6857 (1994).

\bibitem{int} K.~Intriligator, R.G.~Leigh and N.~Seiberg,
  Phys.~Rev.~{\bf D50}, 1092 (1994); \\
  K.~Intriligator, Phys.~Lett.~{\bf B336}, 409 (1994).

\bibitem{PDG} Particle Data Group, Phys. Rev. {\bf D50}, 1173 (1994).

\bibitem{hagiwara} K. Hagiwara, Talk given in YKIS'95 (1995); \\
  K.~Hagiwara, S.~Matsumoto, D.~Haidt and C.S.~Kim, Z. Phys.
  {\bf C64}, 559 (1994).

\bibitem{dimfiveop} N.~Sakai and T.~Yanagida, Nucl. Phys. {\bf B197},
  533 (1982); \\
  S.~Weinberg, Phys. Rev. {\bf D26}, 287 (1982).

\bibitem{hisano2} J.~Hisano, H.~Murayama and T.~Yanagida,
  Nucl. Phys. {\bf B402}, 46 (1993); \\
  R.~Arnowitt and P.~Nath, Phys. Rev. {\bf D49} 1479 (1994).

\end{thebibliography}
\end{document}